\documentstyle[epsf,psfig,rotate,amstex,amssymb]{l-aa} 

\begin{document}

   \thesaurus{08         
              (08.03.2;  
               08.01.1;  
               08.05.3;  
               08.06.3)  
             }

\title{Behaviour of calcium abundance in Am-Fm stars with evolution
\thanks{Based on observations collected at Observatoire de Haute Provence (CNRS),
France, and on data from the ESA HIPPARCOS astrometry satellite.}}

\author{M. K\"unzli \and P. North}

\offprints{P. North}

\institute{Institut d'Astronomie de l'Universit\'e de Lausanne,
 CH-1290 Chavannes-des-Bois, Switzerland
 }
\date{Received 7 July 1997/ Accepted 30 September 1997}

\maketitle

\markboth{M. K\"unzli et al.: Calcium abundance in Am-Fm stars}
{M. K\"unzli et al.: Calcium abundance in Am-Fm stars}

\begin{abstract}
Calcium abundance in the atmosphere of Am stars is examined as a function of
their evolutionary state within the main sequence. New spectroscopic abundances
as well as abundances abtained photometrically by Guthrie (1987) are used,
since they are mutually quite consistent.

The main result of this work is that, contrary to earlier suggestions, calcium
abundance does not tend to be larger in evolved Am stars than in unevolved ones,
for objects distributed along a given evolutionary track in the HR diagram.
The trend appears to be the reverse, if it is present at all.

For our whole sample of Am stars, there is a significant correlation between
calcium abundance and effective temperature, in the sense that the cooler
objects are the most Ca-deficient, hance have the most pronounced Am
peculiarity. This implies an apparent correlation between calcium deficiency
and age, although the lack of Am stars younger than $\log t = 8.6$ seems real.
Our results are fully consistent with the low rate of Am stars observed in
young clusters and with theoretical predictions of time-dependent radiative
diffusion (Alecian 1996).
\end{abstract}

\section{Introduction}

The Am-Fm stars, whose effective temperature lies between 7000 K and 9000 K,
are the coolest chemical peculiar stars on the main sequence (excluding barium
or carbon dwarfs, which owe their peculiarity to binary evolution). Their main 
characteristics are an underabundance of calcium and scandium (about 5 to 10
times lower than in the Sun), a slight overabundance of iron-peak elements, 
a slow rotational velocity ($v\sin i \leq 100 \ \mathrm{km\ s^{-1}}$) and
a high rate of tight binaries (Abt and Levy 1985).

To explain the emergence of chemical anomalies in Am stars, one usually
invokes the radiative diffusion theory developed by Michaud et al. (1983).
This theory predicts that, in a slowly rotating star where the large-scale
meridional circulation is weak enough, helium is no longer
sustained and flows inside the star, gradually disappearing from the atmosphere.
The diffusion process could therefore take place just below the thin H\,{\sc i}
convective zone where the diffusion time is short with respect
to the stellar lifetime; as a first approximation, the chemical elements whose
radiative acceleration is larger than gravity become overabundant
and, in the opposite case, underabundant.

The H\,{\sc i} convective zone becomes deeper as the star evolves on the main
sequence; finally, the c.z. may dredge-up calcium and scandium, leading to 
the normalisation of the surface abundance. Berthet (1992), using data for Am
members of three open clusters, provided some evidence for a trend between 
calcium abundance and evolutionary stage in agreement with the preceding 
scenario; Guthrie (1987) had already suspected such a trend in a sample of field
Am stars. Berthet (1992) proposed an evolutionary scenario for the Am stars
by considering also the $\delta$ Del stars (which have the same abundance
anomalies as the Am stars except for Ca and Sc which are not deficient) and
the metallic A and F giants discovered by Hauck (1986) on the basis of their
enhanced blanketing parameter $\Delta m_2$ of Geneva photometry: an Am star
would evolve into a $\delta$ Del and finally into a metallic F giant, following
a sequence of increasing Ca abundance. However, a comparison of multiplicity
and rotational velocities of metallic A-F giants and of Am stars has shown
that the latter cannot be the progenitors of the former, which casts serious
doubts upon Berthet's scenario (Künzli \& North 1997).

Alecian (1996) has studied theoretically the evolution of calcium abundance in 
the early stages of slowly rotating A and F type stars. His work predicts a 
short phase of calcium overabundance (before $\log t$ = 8) followed by a phase 
of underabundance for some depth values of the mixing zone which is just below 
the H\,{\sc i} convective zone. According to this result, all slowly rotating
A and F stars go through a phase of underabundance of Ca, but only after
$10^8$ years. This prediction is supported by North (1993), who pointed out a 
deficiency of Am stars in young open clusters.

One goal of the present work is to test the results of 
Berthet (1992) and Guthrie (1987). Another is to determine, as far as 
possible, the evolutionary state at which slowly rotating A and F stars become 
Am stars. The knowledge of this  parameter may shed some light on the formation 
of Am stars. To this end, we measured at OHP some known bright Am stars whose Ca
abundance were then determined by optimum fit of synthetic spectra to our
observed ones, taken in the region of the Ca\,{\sc ii} K line. To this sample 
we added the 57 stars of Guthrie (1987) whose Ca abundance was determined from 
the photometric $k$ index and is well-correlated with ours for the eight common 
stars. For each star, we determine the effective temperature from Geneva 
photometry (Künzli et al. 1997) and the absolute magnitude (corrected for 
duplicity) from the Hipparcos parallax. Then, in the HR diagram we compute the 
evolutionary state (defined by our $D_{1000}$ parameter) and the age and mass
by interpolation in the evolutionary tracks of Schaller et al (1992).

\section{The sample and observations}
We have selected  Am stars between 1.5 and 2 solar masses at different distances
from the ZAMS in order to follow the calcium abundance with evolution. The
objects selected come from the catalogue of Hauck \& Curchod (1980) which 
contains 385 Am stars with known spectral type. We have added some metallic 
giant F stars (Hauck 1986) and normal A stars as reference stars. Because of
bad weather, we could observe only 27 Am stars, 2 metallic giant F stars and
2 normal stars.

Our observations were made during two sessions of a few nights' duration each
at Observatoire de Haute-Provence (OHP) in May 1994 and in November 1995, using 
the 1.52m telescope equipped with the AUR\'ELIE spectrograph (Gillet et al. 
1995). The detector is a double-element TH7832 with sets of 2048 photodiodes of 
750 x 13 $\mu$m. We used the grating $N^o$ 2 with 1200 lines/mm; the spectra 
were thus obtained at a reciprocal dispersion of 8 \AA mm$^{-1}$ in the spectral
region centered on the Ca\,{\sc ii} K line [3820\AA, 4035\AA]. Using calibration
spectra of thorium, the reduction was made at Geneva with MIDAS procedures 
for the first mission and at OHP with IHAP procedures for the second mission.
To normalise our spectra, we simply fitted a straight line to the continuum.
For most spectra, a signal-to-noise ratio of 150 was achieved in the continuum.

\section{Calcium abundance}
For the same reasons as those invoked by Berthet (1992), we use the strong
Ca \,{\sc ii} K line to determine the abundance of this element. In the 
atmospheres of Am stars, most of the calcium is ionised due to a low first 
ionisation potentiel, thus the assumption of LTE may fail for the neutral 
calcium lines. However, the wings of the Ca\,{\sc ii} K line are formed deeper 
in the atmosphere than the other lines of ionised elements, making the LTE 
assumption more valid because the density of this ion is higher and the 
transition comes from the ground state. In addition, this line is always
visible even when the star rotates rapidly.

\begin{figure}
\epsfysize=9cm
\leavevmode\epsffile{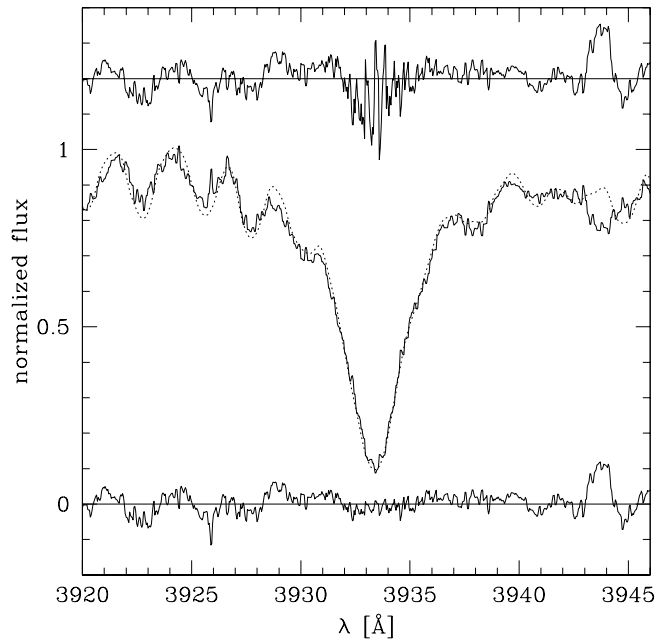}
\caption[]{AUR\'ELIE spectrum of the Am star HD 74190 in the vicinity of 
the Ca\,{\sc ii} K line. The synthetic spectrum (dotted line) with 
$T_{\mathrm{eff}}$ = 7944 K, $\log g$ = 3.84 dex, [M/H] = -0.22 dex,
$\xi_t$ = 3.3 $\mathrm{km\ s^{-1}}$ and $v\sin i$ = 58 $\mathrm{km\ s^{-1}}$
is shown together 
with the observed one (continuous line). The ratio (increased by 0.2 for 
clarity) of the observed and synthetic spectra is shown at the top of the 
figure, while the difference is shown at the bottom. The synthetic spectrum is 
computed with $\log(\frac{N_{\mathrm Ca}}{N_{\mathrm H}}) + 12 = 6.28$.}
\end{figure}

The synthetic spectra are computed using the SYNSPEC code of Hubeny \& Lanz 
(1993) and Kurucz atmosphere models (1995). The Stark, van der Waals and 
radiative widths are those calculated by Kurucz (1989) for iron-peak elements 
lines. The adopted parameters $T_{\mathrm{eff}}$ and $[M/H]$ for the 
atmosphere models come from the colours obtained in the Geneva photometric
system for all stars of the sample, using the calibration by Künzli et al.
(1997). For all Am stars, the surface gravity $\log g$ is deduced from the 
HIPPARCOS parallax and from the apparent $V$ magnitude, while
the mass is obtained by interpolation in the evolutionary tracks of Schaller
et al. (1992). For normal and giant metallic stars, $\log g$ is determined
from Geneva photometry (K\"unzli et al. 1997). The microturbulence $\xi_t$ is 
calculated from Edvardsson et al. (1993) 
or Coupry \& Burkhart (1992) according to the effective temperature and the 
surface gravity of the star. The spectra were then convoluted by a gaussian 
with $FWHM$ = 0.379 \AA\ representing the instrumental profile and by the 
appropriate rotational profile. The $T_{\mathrm{eff}}$, $\log g$, $[M/H]$, 
$\xi_t$ and $v\sin i$ values are listed in Table 1 for the stars observed at 
OHP. The Ca\,{\sc ii} K spectral type given in this table comes from the 
catalogue of Hauck \& Curchod (1980) and the $v\sin i$ from Abt \& Morrell 
(1995) or the Bright Star Catalogue (Hoffleit \& Jaschek 1982). In most
cases, the $v\sin i$ values taken from the literature are completely compatible
with our spectra. But in a few instances, the attempt to fit the observed
spectrum with the synthetic one failed, when the latter was convoluted using
the $v\sin i$ from the literature; in such cases (or when $v\sin i$ was
previously unknown), we used our own estimate, which is quite reliable
because it is based not only on the Ca K line, but also on many metallic
lines spread over the 200 \AA~ wavelength range.

\begin{figure}
\epsfysize=9cm
\leavevmode\epsffile{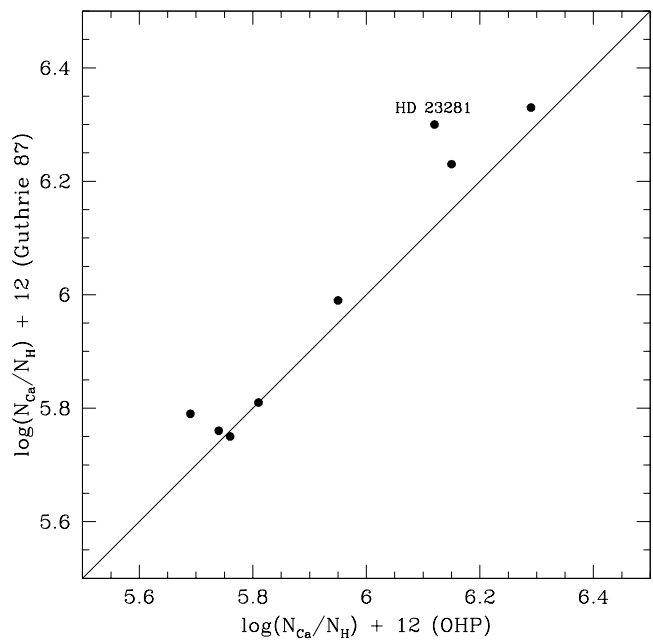}
\caption[]{Comparison between Ca abundances obtained by Guthrie (1987) and those
obtained in this paper. The dispersion of the points from the straight line at 
45 degrees is 0.07 dex.}
\end{figure}

When fitting a spectrum, only two parameters remain free, that is the rotational
velocity and of course the calcium abundance. The effective temperature, 
surface gravity, metallicity and microturbulence are fixed. The 
photometrically determined $T_{\mathrm{eff}}$ is confirmed by the fit of the
Balmer lines H$_\eta$, H$_\zeta$ and H$_\epsilon$ in our spectra, so we did not 
attempt to change its value, although these lines are computed using a 
simplified broadening theory only and not the VCS theory. The calcium abundance 
essentially affects the wings of the K line, because it is saturated. The 
rotational velocity acts on the depth of the line, the wings being only weakly 
affected. The quantitative effects of the effective temperature, rotational 
velocity and calcium abundance are shown on Figures 3, 4 and 5 of Berthet's 
paper (1992) respectively (he used the ADRS code -- see Chmielewski 1979, Lanz 
1987 -- but one sees of course the same behaviour with SYNSPEC). For the above
reasons, we determine the Ca abundance essentially by a fit on the wings and 
then on the depth of the K line. The resulting Ca abundances are given in
Table 1. The internal precision is estimated at about 0.1 dex. As an exemple,
Figure 1 shows the optimal fit of HD 74190 by a synthetic spectrum in the
region of the Ca\,{\sc ii} K line (3933.663 \AA). This fit leads to a relative 
abundance $\log(\frac{N_{\mathrm Ca}}{N_{\mathrm H}}) + 12 = 6.28$. 

Our determinations were tested with the eight stars we have in common
(HD 18557, HD 23281, HD 40062, HD 60652, HD 71297, HD 136403, HD 221675, HD 
223461) with Guthrie (1987), who measured the Ca abundance photometrically by 
the $k$ index. Figure 2 compares the Ca abundances of Guthrie with those
determined at OHP. Except for HD 23281 which shows a difference of 0.18 dex 
between Guthrie's value and ours, there is an excellent correlation. The rms 
scatter around the straight line at 45 degrees is only 0.07 dex. For this 
reason, we include Guthrie's sample in our discussion. Guthrie estimates his 
overall error on Ca abundance at about $\pm 0.3$ dex, but it is probably less.

\begin{table*}
\scriptsize
\hspace{0.35cm}{{\bf Table 1.} Abundances of Ca for the sample of Am-Fm stars.
The uncertainties on $\log t$ are propagated from assumed errors on
$T_{\rm eff}$ and $\log g$ of 270 K and 0.18 dex respectively (Asiain et al. 
1997), which are comfortably large. An asterisk indicates that $M_v$ was
determined not from Hipparcos parallaxes, but from Geneva photometry.}
\vspace{2mm}   
\begin{center}
\begin{tabular}{r r r r r r r c r c r r r r}
\hline\multicolumn{1}{r}{HD}
     &\multicolumn{1}{r}{ST} 
     &\multicolumn{1}{r}{Rem.} 
     &\multicolumn{1}{r}{$T_{\mathrm{eff}} $} 
     &\multicolumn{1}{r}{$\log g$} 
     &\multicolumn{1}{r}{[M/H]} 
     &\multicolumn{1}{c}{$vsin\ i$} 
     &\multicolumn{1}{c}{$\xi_t$} 
     &\multicolumn{1}{r}{$M_v$} 
     &\multicolumn{1}{r}{$\log(\frac{N_{\mathrm Ca}}{N_{\mathrm H}}) + 12$}
     &\multicolumn{1}{c}{$M$} 
     &\multicolumn{1}{c}{$\log t$} 
     &\multicolumn{1}{r}{$D_{1000}$} \\
      \multicolumn{1}{r}{}
     &\multicolumn{1}{r}{Ca\,{\sc ii} K} 
     &\multicolumn{1}{r}{} 
     &\multicolumn{1}{r}{[K]} 
     &\multicolumn{1}{r}{} 
     &\multicolumn{1}{r}{} 
     &\multicolumn{1}{c}{$[\mathrm{km\ s^{-1}}]$} 
     &\multicolumn{1}{c}{$[\mathrm{km\ s^{-1}}]$} 
     &\multicolumn{1}{r}{} 
     &\multicolumn{1}{r}{} 
     &\multicolumn{1}{c}{[M$_{\odot}$]} 
     &\multicolumn{1}{r}{} 
     &\multicolumn{1}{r}{} \\
\hline
861&A2m&SB1&7715&3.90&0.29&35&3.2&1.49&5.95&1.98&8.934 $\pm$ 0.050&1.285\\
2628&A5m&-&7223&3.77&-0.14&20&2.7&1.43&6.15&1.98&8.992 $\pm$	0.081&1.800\\
15385&A5m&V&8154&4.12&0.00&21&3.4&1.85&6.61&1.88&8.793 $\pm$  0.115&0.520\\
17584&F2III&IIIm&6726&3.56&-0.09&149&3.2&$^*$1.26&6.18&1.91&9.112 $\pm$ 0.116
&2.026\\
18557&A2m&V&7614&3.90&0.10&15&3.2&1.57&5.74&1.94&8.958 $\pm$  0.052&1.267\\
21912&A3m&SB1O&8327&4.31&0.00&91&3.3&2.30&6.15&1.80&7.089 $\pm$ 4.000&0.026\\
23281&A3m&-&7689&4.20&-0.02&81&3.2&2.42&6.12&1.68&8.797 $\pm$  0.287&0.199\\
24141&A5m&-&8070&4.22&0.35&54&3.3&2.21&6.38&1.78&8.633 $\pm$	0.405&0.217\\
36484&A2m&SB&8711&4.27&0.00&35&3.2&1.90&5.82&1.93&8.312 $\pm$	0.674	&0.122\\
40062&A5m&-&7030&3.68&0.39&40&2.6&1.32&5.69&2.02&9.001 $\pm$	0.076	&2.110\\
42954&A5m&SB&7384&3.70&0.17&47&3.0&1.10&5.90&2.11&8.934 $\pm$	0.074	&2.087\\
44691&A3m&SB1O&7581&3.76&0.15&21&3.2&1.13&5.90&2.12&8.911$\pm$ 0.064
&1.897\\
60652&A5m&-&7647&3.86&0.44&63&3.2&1.40&6.15&2.01&8.935 $\pm$	0.057&1.455\\
63589&A2m&V&8122&4.25&0.00&35&3.4&2.24&5.85&1.78&8.531 $\pm$	0.603	&0.161\\
67317&A1m&-&7165&4.18&0.43&35&2.7&2.77&5.42&1.55&8.989	$\pm$ 0.229&0.144\\
71297&A5m&-&7712&4.06&-0.03&13&3.2&1.97&6.29&1.81&8.935 $\pm$	0.060&0.698\\
74190&A5m&-&7944&3.84&-0.22&58&3.3&1.13&6.28&2.13&8.868 $\pm$	0.053&1.568&\\
83886&A5m&-&8638&4.18&0.00&100&3.2&1.69&6.27&1.98&8.613 $\pm$	0.220&0.375&\\
84607&F4III&IIIm&7050&3.79&0.09&120&3.0&$^*$1.66&6.46&1.89&9.052 $\pm$ 0.078
&1.642\\
88295&(A0)&normal&8594&4.13&0.00&120&3.3&$^*$1.58&6.36&2.01&8.683 $\pm$ 0.129
&0.515\\
110326&A3m&SBO&7076&4.04&-0.26&65&2.6&2.42&5.54&1.63&9.116 $\pm$0.060&0.592\\
135774&A6m&V&6917&3.80&0.42&35&3.0&1.79&5.78&1.84&9.091 $\pm$	0.078	&1.579\\
136403&A2m&SBO&7670&4.01&0.17&20&3.2&1.87&5.81&1.84&8.955 $\pm$	0.047	&0.844\\
150557&F2III-IV&normal&6849&3.85&-0.18&95&2.7&$^*$2.02&6.20&1.75&9.134$\pm$0.068
&1.323\\
169885&A3m&-&8050&3.91&0.03&60&3.4&1.26&6.11&2.08&8.864	$\pm$ 0.042	&1.306\\
183262&A5m&-&7055&4.06&0.21&73&2.6&2.50&5.90&1.61&9.124	$\pm$ 0.065	&0.499\\
190401&A7m&-&6877&3.81&0.25&40&2.8&1.84&5.86&1.82&9.104 $\pm$	0.077	&1.548\\
193472&A5m&SB2&7123&3.72&0.23&93&2.7&1.38&5.68&1.99&9.010 $\pm$	0.076&1.907\\
213534&A5m&SBO&7632&3.74&-0.31&48&2.7&1.05&6.22&2.15&8.894 $\pm$0.064&1.968\\
221675&A2m&-&7223&3.84&0.53&70&2.8&1.67&5.76&1.89&9.031 $\pm$	0.068	&1.472\\
223461&A2m&-&7879&3.91&0.11&48&3.3&1.38&5.95&2.03&8.899 $\pm$	0.044	&1.282\\
\hline
\end{tabular}
\end{center}
\end{table*}

\begin{figure}
\epsfysize=9cm
\leavevmode\epsffile{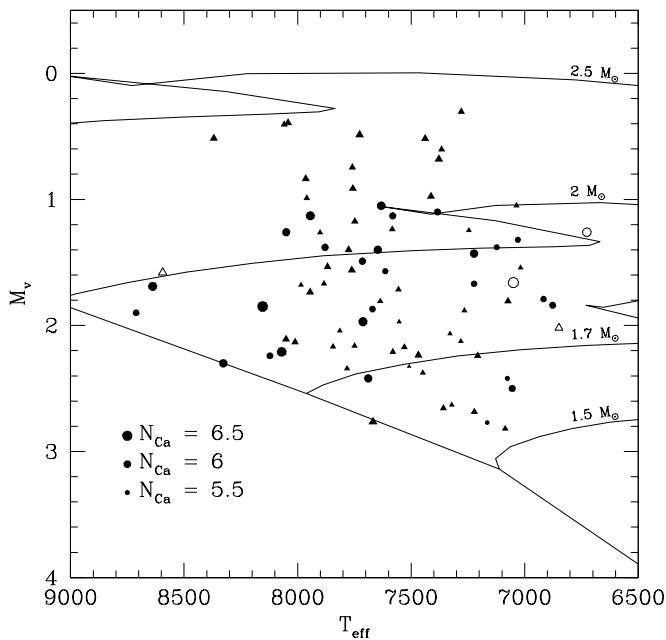}
\caption[]{HR diagram of the 76 Am stars included in the discussion. The 
absolute magnitudes are deduced from Hipparcos parallaxes and corrected for 
duplicity. The effective temperatures are taken from Künzli et al. (1997). The 
stars measured at OHP are represented by black circles and those of Guthrie 
(1987) by black triangles. The other stars measured at OHP  are represented by 
open circles for giant metallic F stars and  by open triangles for normal stars.
The size of the points is related to  the Ca abundance.}
\end{figure}

\section{Discussion}

As part of the discussion will depend on the position of Am stars in the HR 
diagram, it is necessary to present their fundamental
parameters in more details.

The effective temperature is photometrically determined  by the calibration of
the Geneva system (Künzli et al. 1997) which uses both stars with known
fundamental parameters and atmosphere models (Kurucz 1993, 1994, 1996a, 1996b).
The only drawback of the Geneva system, compared to the $uvby \beta$ one, is
its sensitivity to interstellar reddening for A and cooler stars. But, as all
76 Am stars are bright ($m_v \leq$ 7 mag.), IS reddening is negligible. The 
fundamental stars used in this range of temperature are essentially those of 
Blackwell \& Lynas-Gray (1994), who relied on the infrared flux method. These 
authors give an estimated error of about 2\%.

The absolute visual magnitudes are deduced from Hipparcos parallaxes of Am stars
(Proposal 55). For the two metallic F giants and the two normal stars, the 
absolute magnitudes are taken from the calibration of Hauck (1973). For these 
stars, $M_v$ is preceded by an asterisk in Tables 1 and 2. As the stars 
considered are bright, hence near to us, the relative error on the parallax
is small, in general around 7\%, so the error on the absolute magnitude is on 
average 0.15 mag.

As is well known, Am stars are often members of tight binaries, so we have to 
correct $M_v$ for the flux of the companion. For this, we apply the following
correction:
\begin{itemize}
\item[-] For SB2 systems, we compute $\frac{M_1}{M_2} = \frac{K_2}{K_1}$ and 
with the mass-luminosity relation we obtain $\Delta m^* = m_1 - m_{1+2} = 2.5 
\log(1+10^{-\frac{m_1-m_2}{2.5}})$.
\item[-] For SB1 systems, $\Delta m^*$ = 0.2 is assumed, which corresponds to a 
difference of 1.75 magnitudes between the components.
\item[-]If the star has a variable radial velocity according to the BSC or the
catalogue of Renson (1991), $\Delta m^*= 0.2$ mag is also assumed.
\item[-] If the star is only suspected of having a variable radial velocity
according to the BSC (``V?'' remark), no change is made.
\end{itemize}
Indications about multiplicity and variability are given in the column ``Rem.''
of Tables 1 and 2. Except for HD 193472 which is an SB2, all stars are 
marked SB1, V or V? or have no remark.

We also compute a correction for visual binaries with angular separation less
than 5 arcsec. This correction is applied only to HD 42954 which has a companion
of the same magnitude at 0.5 arcsec, and to HD 67317 which has a faint neighbour
at 1.4 arcsec.

The final HR diagram, where these corrections are taken into account, is 
presented in Figure 3.
Stars represented by black circles are those measured at OHP, those represented
by black triangles are from Guthrie (1987). We have also plotted the two
metallic giant F stars (open circles) and one normal star (open triangle). The 
size of each point depends linearly on the logarithmic calcium abundance.
There is a striking deficiency of young Am stars with masses larger than
2 M$_{\odot}$.

\subsection{Behaviour of Ca abundance with $\log t$}
The Barcelona group (Asiain et al. 1997) has kindly transmitted to us
a code which interpolates the age $\log t$ and mass of a star from its effective
temperature and surface gravity, in evolutionary tracks from various authors 
including Schaller et al. (1992). As a first step, $\log g$ was computed from 
the absolute magnitude, assuming an inital mass of 1 $M_\odot$. Then $\log g$ 
was determined by the absolute magnitude and the mass given in the first 
interpolation. This iterative process was stopped as soon as the mass converged 
to $|(\frac{M}{M_\odot})_{i-1}-(\frac{M}{M_\odot})_i| \leq 10^{-3}$. The number 
of iterations seldom exceeded 10. The values of $\log t$ with its internal 
error and of the mass are given in Tables 1 and 2.

When a star is near the ZAMS, the uncertainty on $\log t$ increases
dramatically, because there is a superposition of isochrones in this region. 
Only HD 21912, HD 36484, HD 63589 and HD 204188 are concerned by this problem. 
For them, the error on $\log t$ is larger than 0.5. 
\begin{figure} 
\epsfysize=9cm
\leavevmode\epsffile{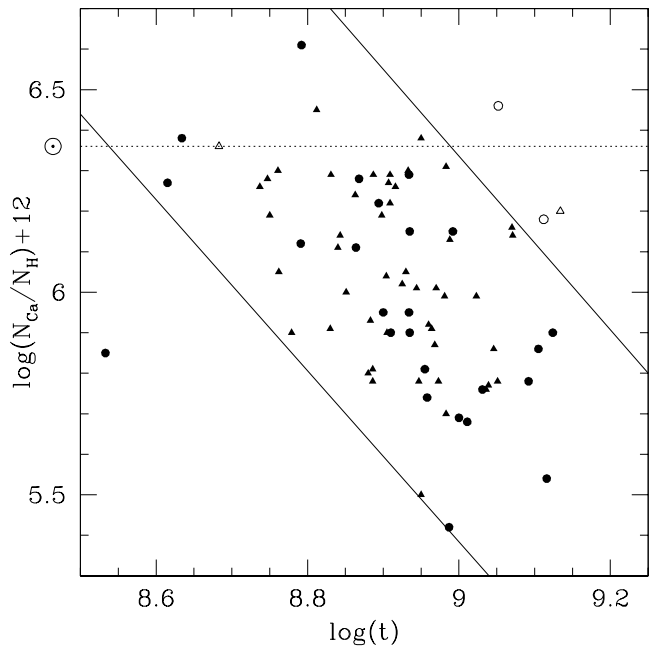}
\caption[]{Calcium abundance as a function of $\log t$. See Figure 3 for the 
key to symbols.}
\end{figure}

\begin{figure}
\epsfysize=9cm
\leavevmode\epsffile{figure5}
\caption[]{Calcium abundance as a function of $T_{\mathrm{eff}}$. See Figure 3 
for the key to symbols.}
\end{figure}

\begin{figure}
\epsfysize=9cm
\leavevmode\epsffile{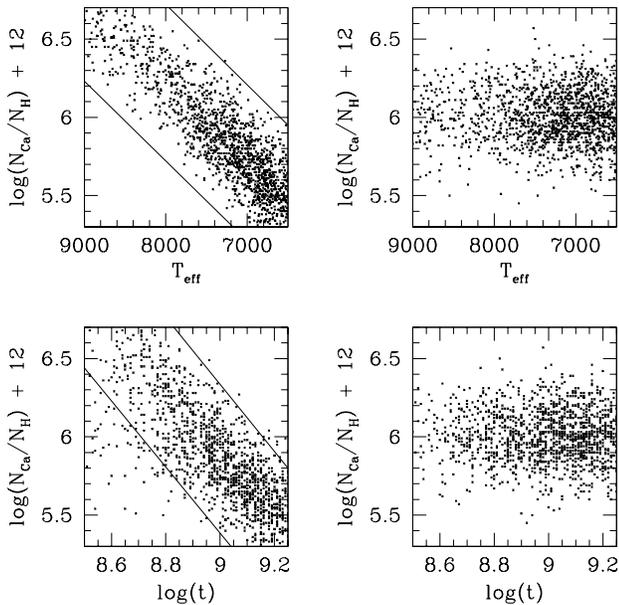}
\caption[]{Diagram of synthesised Am stars. The two leftmost figures reproduce 
exactly what is observed. If there is no relation between Ca abundance and
$T_{\mathrm{eff}}$, there is no relation either between age and Ca abundance,
as shown in both rightmost diagrams. Temperature effects closely mimic age
effects and vice versa, so these two quantities are not independent.}
\end{figure}

Figure 4 shows the behaviour of $\log(\frac{N_{\mathrm Ca}}{N_{\mathrm H}})+12$
with $\log t$.
The age of Am stars spreads between 8.6 and 9.15 with a clear deficiency of Am 
stars in the lower left part of the diagram. It means that A and F stars are 
really deficient in calcium only after $\log t \approx 8.8$. In order to test 
the significance of this result, a $2 \times 2$ contingency table was built, 
using a separation at $\log t =8.875$ and
$\log (\frac{N_{\mathrm Ca}}{N_{\mathrm H}}) + 12 = 5.88$.
We get $\chi ^2 = 9.88$ while the value for the 99.5\%  confidence level 
is 7.88. This dependence between Ca abundance and age seems to confirm
qualitatively Alecian's theory, which predicts a calcium deficiency for slowly 
rotating A and F stars from $\log t = 8$ on, that is 0.6-0.8 dex earlier than 
our observations suggest. However, this relation betrays a correlation 
between Ca abundance and effective temperature (Figure 5): since cooler stars 
are also older, one will necessarily observes a relation between Ca abundance
and age.

We have made some simulations to test the interdependence between age,
effective temperature and Ca abundance. First, one has synthesised a population 
of Am stars with masses and ages distributed at random (assuming a Salpeter 
$IMF$ combined with the relative mass distribution of North 1993); the age
distribution was not uniform, but modified so as to mimic the observed one. By 
interpolation in evolutionary tracks of Schaller et al. (1993), we obtain
both effective temperatures and absolute magnitudes. Finally, we determine
$\log t$ as before and then can plot
$\log(\frac{N_{\mathrm Ca}}{N_{\mathrm H}}) + 12$ vs $\log t$
assuming a relation between effective temperature and Ca abundance. If we impose
the observed relation between $T_{\mathrm{eff}}$ and Ca abundance given in 
Figure 5, we recover the relation between $\log t$ and
$\log(\frac{N_{\mathrm Ca}}{N_{\mathrm H}}) + 12$ shown in Figure 4.
The left part of Figure 6
illustrates this fact, the oblique lines being at the same positions as in 
Figure 5 and 4. On the other hand, if we impose no relation between temperature 
and Ca abundance, there is no relation either between age and Ca abundance 
(Figure 6, right part). These simulations show that temperature effects almost
perfectly mimic age effects and vice-versa, at least when all stars are
considered together. Even if one considers only stars distributed along an
evolutionary path (all having the same mass), it is not possible to discuss age 
effects independently of $T_{\mathrm eff}$ effects, since both parameters are 
intimately related. When a star follows an evolutionary track, the effective
temperature decreases and the age $\log t$ of course increases; if the
correlation between $T_{\mathrm eff}$ and Ca abundance found above holds, then
one may expect that an evolving Am star will become more and more Ca-deficient.
This point is examined in more details in the next Subsection.

\begin{figure}
\epsfysize=9cm
\leavevmode\epsffile{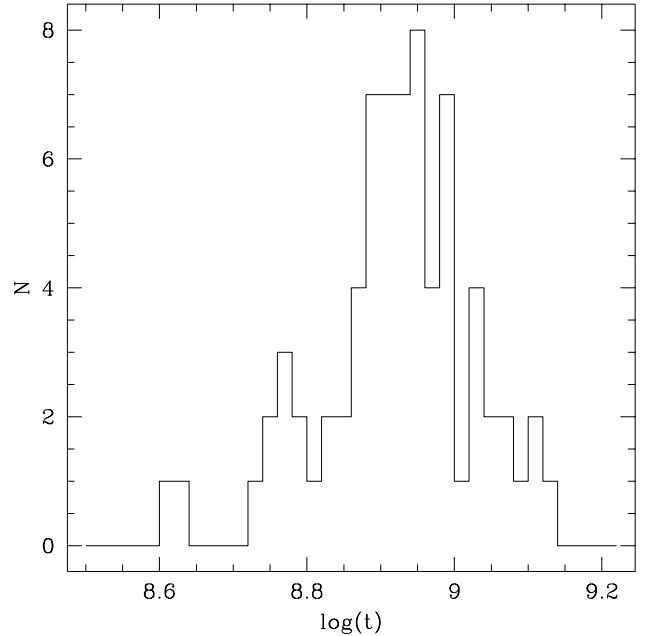}
\caption[]{Histogram of $\log t$ for the Am stars considered in this paper.}
\end{figure}

Even if the effect of age on Ca abundance can not be isolated from that of
$T_{\mathrm eff}$, it is interesting to consider the histogram of ages.
As seen in Figure 7, most of our field Am stars are older than $\log t
\approx 8.8$ (in the histogram, we do not take into account the four stars
having a large error on $\log t$). Although our sample should be considered as
biased because it was defined with the purpose of populating a few evolutionary
tracks as uniformly as possible, Guthrie's sample is not biased towards young
or old objects, and it constitutes two thirds of the whole sample. Therefore,
we think the distribution of Figure 7 shows a real lack of young Am stars.
This fact is coherent with the deficiency of Am stars in young clusters.

\subsection{Behaviour of Ca abundance along an evolutionary track}

\begin{figure}
\epsfysize=9cm
\leavevmode\epsffile{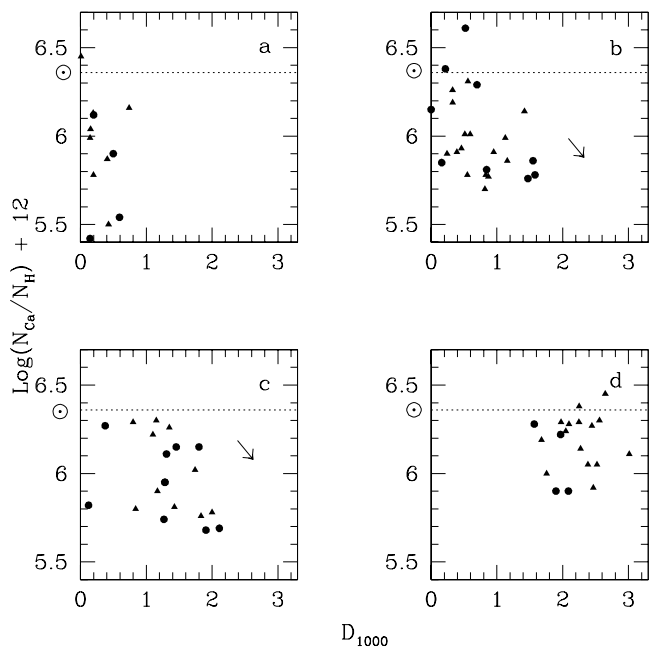}
\caption[]{$\log(\frac{N_{\mathrm Ca}}{N_{\mathrm H}}) + 12$ as a function of 
$D_{1000}$ for different range of mass;
a: 1.5 $\leq M/{\mathrm M_{\odot}} \leq$ 1.7,
b: 1.7 $\leq M/{\mathrm M_{\odot}} \leq$ 1.9,
c: 1.9 $\leq M/{\mathrm M_{\odot}} \leq$ 2.1,
d: $ M/{\mathrm M_{\odot}} \geq$ 2.1.
See Figure 3 for the signification of symbols.}
\end{figure}

In order to know how the Ca abundance tends to vary with evolution for stars
{\it of a given mass}, we define the $D_{1000}$ parameter as the difference
between the present effective temperature of the star and the one it had on the 
ZAMS, divided by 1000 just to have a number close to unity. In
other words, this parameter represents the ``horizontal'' component of the 
distance covered by the star in the HR diagram along an evolutionary track: 
\begin{equation}
D_{1000} = \frac   {T_{\mathrm eff}-T_{\mathrm eff}({\mathrm ZAMS})}
   {1000} \nonumber
\end{equation}
This parameter therefore measures a kind of age, i.e.~the time which has
expired since the star left the zero-age main sequence, although it is of
course not on a linear scale. 

$D_{1000}$ is correlated with the depth of the convective zone
and consequently with Ca abundance, according to
the scenario of Berthet (1992). To show this dependence, we have defined four
mass ranges: 1.5 $\leq M/{\mathrm M_{\odot}} \leq$ 
1.7, 1.7 $\leq M/{\mathrm M_{\odot}} \leq$ 1.9,
1.9 $\leq M/{\mathrm M_{\odot}} \leq$ 2.1 and
$M/{\mathrm M_{\odot}} \geq$ 2.1.
The Ca abundance was then plotted against $D_{1000}$
for each of these ranges (Figure 8). If Berthet's (1992) idea was right, we 
should see an increasing calcium abundance with increasing $D_{1000}$ because
at a constant mass, $D_{1000}$ increases along a given evolutionary track.
Figure 8 shows that this is definitely not the case: there is no correlation
between $\log(\frac{N_{\mathrm Ca}}{N_{\mathrm H}})$ and $D_{1000}$. In Figure 
8b and 8c, we rather see the reverse correlation, i.e.~an increasing calcium 
deficiency with evolution. This rough correlation is in agreement with the 
preceding result: $T_{\mathrm eff}$ decreases with evolution and Ca gets more 
depleted. 

If we superimpose Figure 8a and 8d, a positive correlation appears between 
$\log(\frac{N_{\mathrm Ca}}{N_{\mathrm H}})$ and $D_{1000}$, which is due to a 
temperature effect 
but not to any evolution effects. Guthrie (1987) analysed the behaviour of 
$\log(\frac{N_{\mathrm Ca}}{N_{\mathrm H}})$ with $\delta c_o'$
(which is a photometric parameter
in the Str\"omgren system similar to $D_{1000}$) without any distinction of
mass or temperature, and Berthet (1992) did not take into account the effect 
of temperature either; this is probably the main reason for the difference
between their result and ours.

\begin{table*}
\scriptsize
\hspace{2.5cm}{{\bf Table 2.} Parameters of age, masse, temperature and
absolute magnitude for the stars of Guthrie (1987).}
\vspace{2mm}   
\begin{center}
\begin{tabular}{r r r r r c  r r r}
\hline\multicolumn{1}{r}{HR}
     &\multicolumn{1}{r}{HD}
     &\multicolumn{1}{r}{Rem.}
     &\multicolumn{1}{r}{$T_{\mathrm{eff}} $}
     &\multicolumn{1}{r}{$M_v$}
     &\multicolumn{1}{r}{$\log(\frac{N_{\mathrm Ca}}{N_{\mathrm H}}) + 12$}
     &\multicolumn{1}{c}{$M$}
     &\multicolumn{1}{c}{$\log(\mathrm{t})$} 
     &\multicolumn{1}{r}{$D_{1000}$} \\
      \multicolumn{1}{r}{}
     &\multicolumn{1}{r}{}
     &\multicolumn{1}{r}{}
     &\multicolumn{1}{r}{[K]}
     &\multicolumn{1}{r}{}
     &\multicolumn{1}{r}{}
     &\multicolumn{1}{c}{[${\mathrm M_{\odot}}$]}
     &\multicolumn{1}{c}{} 
     &\multicolumn{1}{r}{} \\
\hline
 178&   3883&  -       &  7279 &0.30 &6.11& 2.34 & 8.851 $\pm$   0.114&  3.011\\
 290&   6116&  -       &  7964 &0.84 &6.29& 2.25 & 8.831 $\pm$   0.062&  1.976\\
 418&   8801&  -       &  7222 &2.69 &6.13& 1.57 & 8.988 $\pm$   0.193&  0.192\\
    540&  11408&  -       &  8051 &2.11 &6.26& 1.80 & 8.736 $\pm$0.240&  0.325\\
    976&  20210&  SBO     &  7509 &2.32 &5.50& 1.70 & 8.951 $\pm$0.111&  0.427\\
    984&  20320&  SBO     &  7530 &2.17 &6.01& 1.74 & 8.970 $\pm$0.073&  0.594\\
   1248&  25425&  V?&  8059 &0.41 &6.05& 2.45 & 8.762 $\pm$   0.072&  2.522\\
   1368&  27628&  SB1O    &  7086 &2.82 &5.99& 1.53 & 9.023 $\pm$0.215&  0.146\\
   1414&  28355&  - &  7761 &1.56 &6.30& 1.95 & 8.933 $\pm$   0.046&  1.152\\
   1428&  28546&  V?&  7468 &2.24 &6.31& 1.71 & 8.982 $\pm$   0.077&  0.561\\
   1511&  30121&  - &  7636 &1.81 &5.91& 1.86 & 8.964 $\pm$   0.043&  0.951\\
   1519&  30210&  SB1?    &  7960 &0.99 &6.00& 2.19 & 8.851 $\pm$0.172&  1.755\\
   1528&  30453&  SB1O    &  7413 &0.98 &6.29& 2.17 & 8.909 $\pm$0.074&  2.248\\
   1627&  32428&  - &  7246 &1.24 &5.78& 2.05 & 8.972 $\pm$   0.075&  2.000\\
   1670&  33204&  - &  7207 &2.24 &6.16& 1.70 & 9.070 $\pm$   0.053&  0.738\\
   1672&  33254&  SBO     &  7553 &1.97 &5.70& 1.80 & 8.983 $\pm$0.048&  0.819\\
   1689&  33641&  V &  7868 &1.53 &6.22& 1.97 & 8.909 $\pm$   0.043&  1.100\\
   2566&  50643&  - &  8042 &0.39 &6.30& 2.46 & 8.761 $\pm$   0.072&  2.558\\
   3320&  71267&  - &  7757 &0.91 &6.24& 2.21 & 8.862 $\pm$   0.072&  2.046\\
   3354&  72037&  - &  7814 &2.04 &5.78& 1.80 & 8.887 $\pm$   0.098&  0.550\\
   3619&  78209&  - &  7281 &2.13 &5.78& 1.73 & 9.051 $\pm$   0.045&  0.831\\
   3655&  79193&  SBO     &  7556 &1.71 &5.99& 1.89 & 8.981 $\pm$0.047&  1.128\\
   3988&  88182&  - &  7945 &1.74 &6.29& 1.90 & 8.886 $\pm$   0.053&  0.800\\
   4021&  88849&  - &  7074 &1.81 &6.14& 1.83 & 9.072 $\pm$   0.067&  1.418\\
   4237&  93903&  SB1O    &  7759 &0.75 &6.14& 2.28 & 8.844 $\pm$0.073&  2.271\\
   4286&  95256&  - &  7985 &1.68 &5.80& 1.93 & 8.880 $\pm$   0.050&  0.837\\
   4424&  99859&  - &  8011 &2.13 &6.19& 1.79 & 8.750 $\pm$   0.234&  0.326\\
   4454& 100518&  - &  7901 &1.26 &5.81& 2.07 & 8.886 $\pm$   0.049&  1.427\\
   4535& 102660&  - &  7329 &2.07 &5.77& 1.76 & 9.039 $\pm$   0.045&  0.871\\
   4543& 102910&  - &  7358 &2.66 &6.04& 1.60 & 8.902 $\pm$   0.277&  0.152\\
   4650& 106251&  - &  7448 &2.37 &5.87& 1.68 & 8.969 $\pm$   0.110&  0.406\\
   4750& 108642&  SB1O &  7884 &1.67 &5.90& 1.92 & 8.905 $\pm$   0.044&  1.170\\
   4751& 108651&  SBO &  7782 &2.34 &5.90& 1.71 & 8.780 $\pm$   0.276&  0.243\\
   4847& 110951&  SBO &  7036 &1.05 &5.92& 2.12 & 8.960 $\pm$   0.115&  2.466\\
   4866& 111421&  - &  7727 &0.49 &6.45& 2.39 & 8.813 $\pm$   0.075&  2.648\\
   5045& 116303&  - &  7366 &0.60 &6.05& 2.20 & 8.929 $\pm$   0.109&  2.839\\
   5405& 126661&  - &  7378 &0.68 &6.38& 2.16 & 8.950 $\pm$   0.108&  2.283\\
   5752& 138213&  SBO &  8369 &0.52 &6.28& 2.41 & 8.747 $\pm$   0.061&  2.096\\
   5845& 140232&  - &  7844 &2.17 &5.91& 1.77 & 8.829 $\pm$   0.166&  0.392\\
   6555& 159560&  SBO  &  7321 &2.63 &5.78& 1.60 & 8.947 $\pm$   0.211&  0.197\\
   6813& 166960&  V &  7265 &1.88 &5.86& 1.82 & 9.046 $\pm$   0.052&  1.156\\
   7019& 172741&  - &  7438 &0.52 &6.27& 2.24 & 8.908 $\pm$   0.109&  2.445\\
   7056& 173648&  SB1O &  7748 &1.17 &6.19& 2.10 & 8.898 $\pm$   0.058&  1.679\\
   7833& 195217&  - &  7581 &2.21 &6.01& 1.73 & 8.944 $\pm$   0.096&  0.514\\
   7990& 198743&  SBO &  7018 &1.54 &5.76& 1.93 & 9.035 $\pm$   0.089&  1.832\\
   8210& 204188&  SBO &  7668 &2.76 &6.45& 1.63 & 7.283 $\pm$   4.000&  0.008\\
   8410& 209625&  SB1O&  7583 &1.24 &6.02& 2.07 & 8.925 $\pm$   0.061&  1.742\\
   8970& 222377&  - &  7749 &2.16 &5.93& 1.76 & 8.884 $\pm$   0.122&  0.459\\
   9025& 223461&  - &  7879 &1.39 &6.26& 2.02 & 8.900 $\pm$   0.044&  1.351\\
\hline
\end{tabular}
\end{center}
\end{table*}

\section{Conclusion}

We have analysed the dependence of Ca abundance with evolution in the
atmosphere of Am stars. In this paper, we put forward the impossibility of 
isolating the dependence of Ca abundance on time alone, because when a star 
evolves, time is intimately related to temperature so that these two parameters 
cannot be discussed separately.

Cool and old Am stars are more Ca-deficient than hotter and younger ones. So, 
when an Am star evolves its Ca abundance probably decreases. But we can only
guess that the physical cause for this is the decreasing temperature in the
star's envelope just below the H\,{\sc i} convective zone, which implies a 
decrease of the radiative acceleration applied to the Ca\,{\sc ii} ions. 
However, the parallel decrease of the surface gravity acts in the opposite
sense and only detailed, time-dependent modelling such as Alecian's will allow
a proper interpretation of the data. 

The $D_{1000}$ parameter, which measures the evolutionary state of a star, 
depends on both age and temperature. For constant mass, $D_{1000}$ follows an 
evolutionary track and so measures directly the effect of evolution. We have
shown that there is no positive correlation between $D_{1000}$ and Ca abundance,
contrary to the claims of Berthet (1992) and Guthrie (1987). There is rather a 
decrease of the calcium abundance as evolution proceeds. In their study,
Berthet (1992) and Guthrie (1987) have mixed stars with different masses and
temperatures, which probably led them to a wrong conclusion, essentially due to 
a temperature effect.

Most ages of our field Am stars spread between $\log t$ = 8.6 and 9.2, which 
agrees well with the lack of Am stars in young open clusters (North 1993).

\acknowledgements{We thank Dr. Simon Jeffery for having transmitted to us a
file with the continous metallic opacities to be used with the SYNSPEC code,
and Mrs Barbara Wilhelm for the correction of the English text.
Constructive criticism by the referee is also gratefully acknowledged.
This paper received the support of the Swiss National Science Foundation.}

\end{document}